\pdfoutput=1

\documentclass[aps,pra, preprint, tightenlines, superscriptaddress]{revtex4-1}
\usepackage{graphicx}  % needed for figures
\usepackage{dcolumn}   % needed for some tables
\usepackage{bm}        % for math
\usepackage{amssymb}   % for math
\usepackage{amsmath}

\usepackage{setspace}
\usepackage[small]{titlesec}
\begin{document}

\title{Valley Polarization in Size-Tunable Monolayer Semiconductor Quantum Dots}

\author{Guohua Wei}
\affiliation{Applied Physics Program, Northwestern University, 2145 Sheridan Road, Evanston, IL 60208, USA}

\author{David A. Czaplewski}
\affiliation{Center for Nanoscale Materials, Argonne National Laboratory, 9700 S Cass Avenue, Argonne, IL 60439, USA}
\author{Erik J. Lenferink}
\affiliation{Department of Physics and Astronomy, Northwestern University, 2145 Sheridan Road, Evanston, IL 60208, USA\vspace{3em}}
\author{Teodor K. Stanev}
\affiliation{Department of Physics and Astronomy, Northwestern University, 2145 Sheridan Road, Evanston, IL 60208, USA\vspace{3em}}
\author{Il Woong Jung}
\affiliation{Center for Nanoscale Materials, Argonne National Laboratory, 9700 S Cass Avenue, Argonne, IL 60439, USA}
\author{Nathaniel P. Stern}\email{n-stern@northwestern.edu}
\affiliation{Applied Physics Program, Northwestern University, 2145 Sheridan Road, Evanston, IL 60208, USA}
\affiliation{Department of Physics and Astronomy, Northwestern University, 2145 Sheridan Road, Evanston, IL 60208, USA\vspace{3em}}

\begin{abstract}
\setstretch{1.25}
Three-dimensional confinement allows semiconductor quantum dots (QDs) to exhibit size-tunable electronic and optical properties that enable a wide range of opto-electronic applications from displays, solar cells and bio-medical imaging to single-electron devices. Additional modalities such as spin and valley properties can provide further degrees of freedom requisite for quantum information and spintronics. When seeking to combine these material features into QD structures, however, confinement can cause hybridization that inhibits the robustness of these emergent properties for insertion into quantum devices. Here, we show that a new class of laterally-confined materials, monolayer MoS$_2$ QDs, can be created through top-down nanopatterning of an atomically-thin two-dimensional semiconductor so that they exhibit the same valley polarization as in a continuous monolayer sheet. Semiconductor-compatible nanofabrication process allows for these low-dimensional materials to be integrated into complex systems, an important feature for advancing quantum information applications.  The inherited bulk spin and valley properties, the size dependence of excitonic energies, and the ability to fabricate MoS$_2$ QDs using semiconductor-compatible processing suggest that monolayer semiconductor QDs have the potential to be multimodal building blocks of integrated quantum information and spintronics systems.

\end{abstract}

%\pacs{}
\maketitle

Semiconductor quantum dots (QDs) can be tailored to exhibit unique optical properties important for diverse opto-electronic applications ranging from light emitting devices~\cite{sun2007bright}, energy harvesting technologies~\cite{kramer2013architecture}, and medical therapies~\cite{azzazy2007diagnostics}, to enabling rich fundamental scientific advances in low-dimensional spintronics and quantum information processing~\cite{trauzettel2007spin,loss1998quantum,petta2005coherent}.
The recent emergence of atomically-thin two-dimensional (2D) layered materials such as graphene and transition metal dichalcogenides (TMDs) provides a rapidly-growing class of low-dimensional materials that exhibit strong carrier confinement in one dimension while preserving a bulk-like dispersion in the 2D plane. As with bulk materials, lateral confinement of these 2D materials such as in graphene QDs~\cite{ponomarenko2008chaotic,trauzettel2007spin} enables size-dependent control of optical and electronic properties in flat atomic-scale crystals layers.

\begin{figure*}
\includegraphics[scale=1]{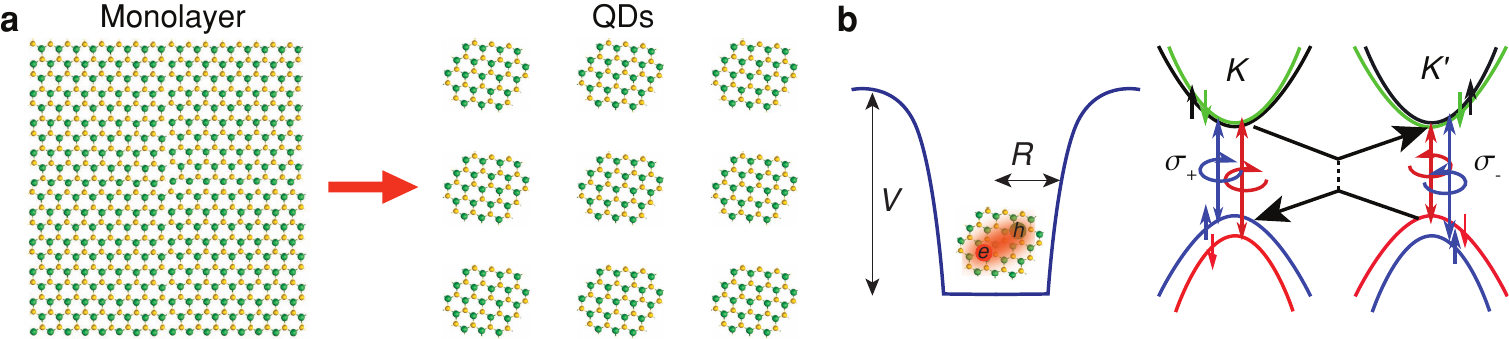}
\caption{\textbf{Monolayer TMD quantum dots.} \textbf{a}, Patterning of a 2D monolayer semiconductor into QDs with lateral confinement. \textbf{b}, 2D quantum confinement of excitons in a potential well of depth $V$ with radius $R$. Inversion asymmetry in monolayer MoS$_{2}$ gives rise to valley-specific coupling to circularly-polarized light. Quantum confinement can potentially enhance intervalley scattering (black lines) and suppress valley polarization.}
\label{fig:MOSQD}
\end{figure*}

Distinct from traditional semiconductors, 2D TMDs exhibit additional layer-dependent properties originating from their crystal symmetries. Of particular recent interest, the spatial inversion asymmetry of the hexagonal crystal structure of monolayer TMD semiconductors combined with large spin-orbit coupling creates degenerate, but distinct, valleys in the band structure with opposite Berry curvature~\cite{xiao2012coupled}. Monolayer TMDs thus support several internal electronic degrees of freedom in the spin and valley pseudospin~\cite{xiao2012coupled,xu2014spin}. The coupled spin-valley dynamics have been extensively studied both optically and electrically~\cite{zeng2012valley,mak2012control,mak2014valley}, with control over the electronic spin and valley pseudospin demonstrating potential for information processing in atomically-thin materials. As observed in silicon and graphene QDs~\cite{goswami2007controllable,saraiva2009physical,trauzettel2007spin}, quantum confinement can cause intervalley coupling and valley hybridization that render the valley pseudospin to no longer be a good quantum number (Fig.~\ref{fig:MOSQD}b).  As chemical, structural, electronic, and optical control of 2D semiconductors continues to evolve, and their lateral dimensionality is further reduced~\cite{gopalakrishnan2015electrochemical}, an open fundamental question is whether the valley polarization of excitons is robust to electronic confinement in monolayer TMD QDs (Fig.~\ref{fig:MOSQD}a).

Here, we show that a new class of laterally-confined semiconductor materials can be created through nanopatterning of single TMD layers into monolayer QDs that maintain the valley properties lost with quantum confinement in other materials such as graphene and silicon. The controlled semiconductor-compatible nanofabrication of monolayer semiconductor QDs enables investigation of valley polarization in a regime of low-dimensional quantum confinement with the possibility for integration into more complex devices for functional applications. Our measurements show that monolayer TMD QDs in the weak confinement regime inherit the valley-selective band structure of the continuous monolayer while exhibiting controlled optical properties that depend on their lateral size. The observation of negligible valley hybridization with quantum confinement suggests that monolayer TMD QDs are a promising platform for exploring the valley degrees of freedom of a single low-dimensional electron.

%\section*{Results}
\section*{Size-dependent quantum confinement in 2D monolayer semiconductor QDs}

Conventional semiconductor QDs are typically obtained via chemical synthesis of nanocrystals~\cite{dabbousi1997cdse}, epitaxial heterostructure growth~\cite{petroff1994mbe}, or electrical definition using patterned electrodes~\cite{hanson2007spins}.  Lateral quantum confinement of the tightly-bound excitons in a 2D monolayer TMD requires a nanometer-scale electronic potential well. Control of confinement in monolayers at this scale by patterning electrodes or chemical synthesis is difficult and not precise. Monolayer TMD nano-flakes with diameter 2-9~nm can be prepared by liquid exfoliation and chemical processes~\cite{gan2015quantum,gopalakrishnan2015electrochemical}, but isolation of  monolayer QDs from the solution for integration into functional devices is challenging and the size of the QDs is poorly controlled.  Recently, quantum emitter behavior in 2D TMDs has been demonstrated from defect states in monolayer WSe$_2$~\cite{he2015single,srivastava2015optically,koperski2015single,chakraborty2015voltage}. Although interesting for quantum information applications, these randomly distributed localized states do not exhibit size-dependent electron properties indicative of quantum dot confinement or the valley polarization characteristic of excitons in the monolayer TMD band structure. Creating monolayer QDs using controllable semiconductor-compatible processing would allow for systematic exploration of the valley degree of freedom of layered TMD materials in a confined regime and for integration of these low-dimensional materials into more complex functional systems.

\begin{figure*}
\includegraphics{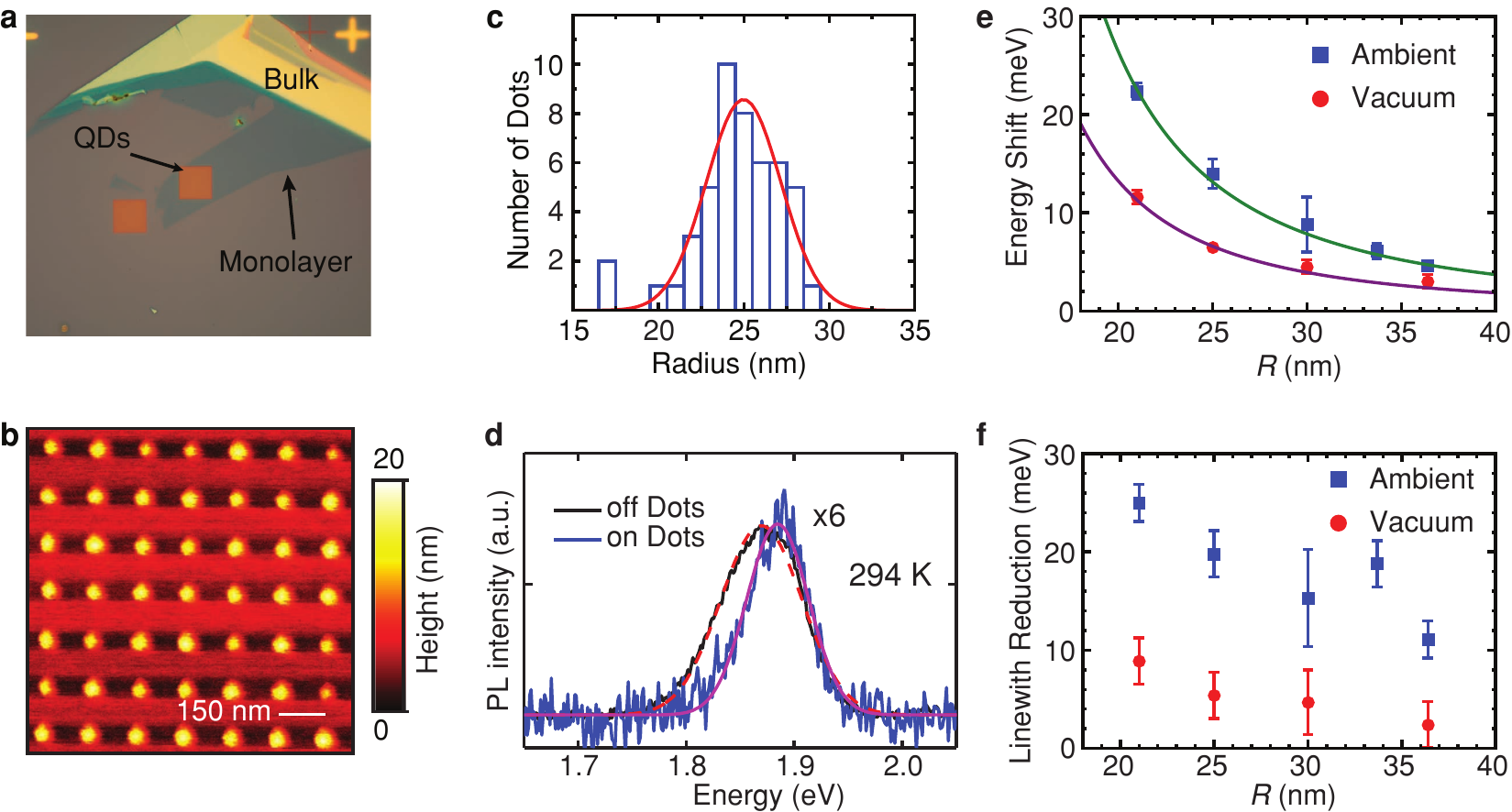}
\caption{\textbf{Characterization of monolayer MoS$_{2}$ quantum dots.} \textbf{a}, Optical image of a sample. A 12 $\mu$m $\times$ 12 $\mu$m square of the monolayer MoS$_2$ flake is processed into laterally confined QDs. \textbf{b}, AFM scan of a $1$ $\mu$m $\times 1$ $\mu$m region of patterned QDs. The dot spacing is 150 nm. The top layer of resist is not removed after processing to protect the QDs.  \textbf{c}, QD size distribution from the AFM scan in \textbf{a}. The size distribution is characterized by a normal distribution (red). For the sample shown, the average QD radius is $R = 25$~nm with a distribution standard deviation of 3~nm. \textbf{d}, PL spectra of QDs (on dots) and of a continuous monolayer (off dots) from a single monolayer flake are shown in blue and black, respectively. Each spectrum is fit with a Gaussian function. \textbf{e}, QD size dependence of the exciton energy shift measured in ambient (blue square) and vacuum (red circle) conditions at room temperature. \textbf{f}, Linewidth reduction of QDs measured in ambient (blue square) and in vacuum (red circle) conditions at room temperature.}
\label{fig:EnergyShift}
\end{figure*}

We develop a top-down method to fabricate size-controlled monolayer MoS$_2$ QDs using electron beam (e-beam) lithography. Semiconductor processing allows for highly accurate registration of TMD QD patterns with respect to other features at the 10~nm length scale, while size control is obtained by etching conditions. Since direct exposure of monolayer MoS$_{2}$ to an electron beam degrades the material’s optical quality and electrons with energy greater than 80~keV can break Mo-S bonds and introduce structural defects~\cite{garcia2014analysis}, patterning methods requiring high e-beam exposure doses typical in high resolution e-beam lithography~\cite{grigorescu2007sub} are not acceptable. We overcome this limitation by only exposing the surrounding area of each dot followed by a reactive ion etch to further reduce the monolayer size to the desired lateral scale (for more information, see Supplementary Information). Unlike the localized excitons~\cite{he2015single,srivastava2015optically,koperski2015single,chakraborty2015voltage} or chemically prepared TMD  nano-flakes~\cite{gan2015quantum,gopalakrishnan2015electrochemical}, our top-down fabrication process allows control of the position and of the lateral size of the delocalized exciton in a TMD QD. For each sample, a partial region of a continuous monolayer flake is patterned so that control measurements can be performed on a single monolayer both on and off the patterned region (Fig.~\ref{fig:EnergyShift}a). Fig.~\ref{fig:EnergyShift}b shows an atomic force microscope (AFM) image from one sample with its QD size distribution shown in Fig.~\ref{fig:EnergyShift}c. The QD sizes are approximately normally distributed. Samples with an average QD radius as small as 20~nm were fabricated.

The effect of the lateral confinement on the MoS$_2$ excitons is measured by photoluminescence (PL) spectroscopy with an approximate spot size of 1~$\mu$m. A representative PL spectrum from a sample with average QD radius $R = 25$~nm (Fig.~\ref{fig:EnergyShift}c) is shown in Fig.~\ref{fig:EnergyShift}d. Compared to the un-patterned monolayer PL spectrum, the peak PL energy of the QDs made from the same monolayer flake blue shifts and the linewidth (full width at half maximum of a Gaussian lineshape) narrows.   A size-dependent energy shift of the QD PL relative to the continuous monolayer emission is evident~(Fig.~\ref{fig:EnergyShift}e). The energy shift is consistently smaller when measured in a vacuum chamber with pressure lower than 200~mTorr compared to in ambient conditions. The larger energy shift in air can be caused by adsorption of water molecules and oxygen~\cite{late2012hysteresis}, which can modify the monolayer electronic properties such as effective mass~\cite{ramasubramaniam2012large} and reduce the effective size of the QD. Similar increased confinement and reduced effective QD size have been observed in semiconductor nanocrystal QDs while exposed to air~\cite{cordero2000photo,van2001photooxidation,sykora2010effect}.

Excitons in a weak confinement regime can be modeled using an envelope function approach to estimate the modified eigenenergies of the exciton center-of-mass wavefunction. The nearly-circular monolayer QDs are modeled as a two-dimensional axisymmetric infinite potential of radius $R$, which predicts a $1/R^{2}$ dependence for the ground state exciton confinement energy $E_{\textrm{ex}}$~\cite{bera2008energy,woggon1997optical}:
\begin{equation}
E_{\textrm{ex}} = E_{0} + \frac{\hbar^2x_{0}^2}{2\mu_\textrm{ex}R^{2}}
\label{eq:energy}
\end{equation}
Here, E$_{0}$ is the exciton energy without in-plane quantum confinement, $\hbar$ is the reduced Planck's constant, $x_{0}\simeq 2.4048$ is the first root of the zero-order Bessel function,  $\mu_\textrm{ex}$ is the reduced mass of the exciton ( $\mu_\textrm{ex} = m_{e}*m_{h}/(m_{e}+m_{h})$). In the weak confinement regime $R\gg a_{B}$ ($a_{B}$ is the exciton Bohr radius), the envelope function approach is a good approximation when the exciton has large binding energy~\cite{woggon1997optical}. This is the case for monolayer MoS$_{2}$, in which $a_{B} \sim 1$~nm is small compared to the patterned QD radius in our experiment and the exciton binding energy is approximately  0.9~eV~\cite{cheiwchanchamnangij2012quasiparticle,ramasubramaniam2012large}. Since the QD size measured by AFM is a convolution of the AFM tip and dot geometry, and the reactive etching process or environment can alter the electronic properties of the monolayer edge, the effective QD radius can be several nanometers smaller than the measured size (see Supplementary Information).  The energy shift $\Delta E = E_{\textrm{ex}} -  E_{0}$, measured both in ambient and vacuum conditions, is fit to a modified form of equation~(\ref{eq:energy}) with  $R \rightarrow R -\varDelta R$, where $R-\varDelta R$ is the effective QD radius. Measurements in both ambient and vacuum conditions show a shift of effective QD radius $\varDelta R \sim 8$~nm. The exciton effective mass $\mu_{\textrm{ex}}$ from the measurement in vacuum (ambient) is extracted from the fit to be $0.116 \pm 0.032$ $m_{0}$~($0.058 \pm 0.006$ $m_{0}$), where $m_{0}$ is the free electron mass. These results are fairly close to the theoretically predicted exciton effective mass of $\sim~0.19m_{0}$ ~\cite{cheiwchanchamnangij2012quasiparticle,peelaers2012effects}, especially for the measurement in vacuum. Since the effective mass of excitons depends on the dielectric environment~\cite{ramasubramaniam2012large,woggon1997optical}, a difference between the fit effective mass and the theoretical calculation for a monolayer is expected for our samples, which are sandwiched between the SiO$_{2}$ substrate and the e-beam resist. The difference between the ambient and vacuum parameters suggests that the QDs are influenced by adsorption from the environment.

The exciton photoluminescence linewidth $\Gamma$ decreases with confinement compared to the unpatterned MoS$_2$ monolayer as shown in Fig.~\ref{fig:EnergyShift}f.  The exciton linewidth is reduced more for smaller QDs, with the narrowest linewidth measured to be about 25~meV smaller compared to a typical MoS$_2$ monolayer linewidth ($\Gamma \sim 90$~meV). The linewidth reduction measured in vacuum is smaller compared to that in ambient conditions, but the exciton linewidth is still narrower than that of the unpatterned continuous monolayer. The PL linewidth of the QD ensemble is determined by two competing mechanisms. Quantum confinement narrows the homogeneous linewidth~\cite{zhao2002effect} whereas the inhomogeneous QD size distribution broadens the linewidth. The width of the QD size distributions (Fig.~\ref{fig:EnergyShift}b) is about 10\%, which introduces a 20\% fluctuation in confinement energy due to the $1/R^{2}$ energy dependence. Because of the small energy shifts in the weak confinement regime, the size distribution only introduces less than 2~meV of inhomogeneous broadening. The narrowing of the PL linewidth implies that the quantum confinement effect dominates the inhomogeneous broadening introduced by the fabrication process. Although the difference between vacuum and ambient conditions deserves further investigation, the $1/R^{2}$ dependence of the exciton energy shift and narrower PL linewidth demonstrate lateral quantum confinement of excitons in a TMD monolayer.

\begin{figure*}
\includegraphics{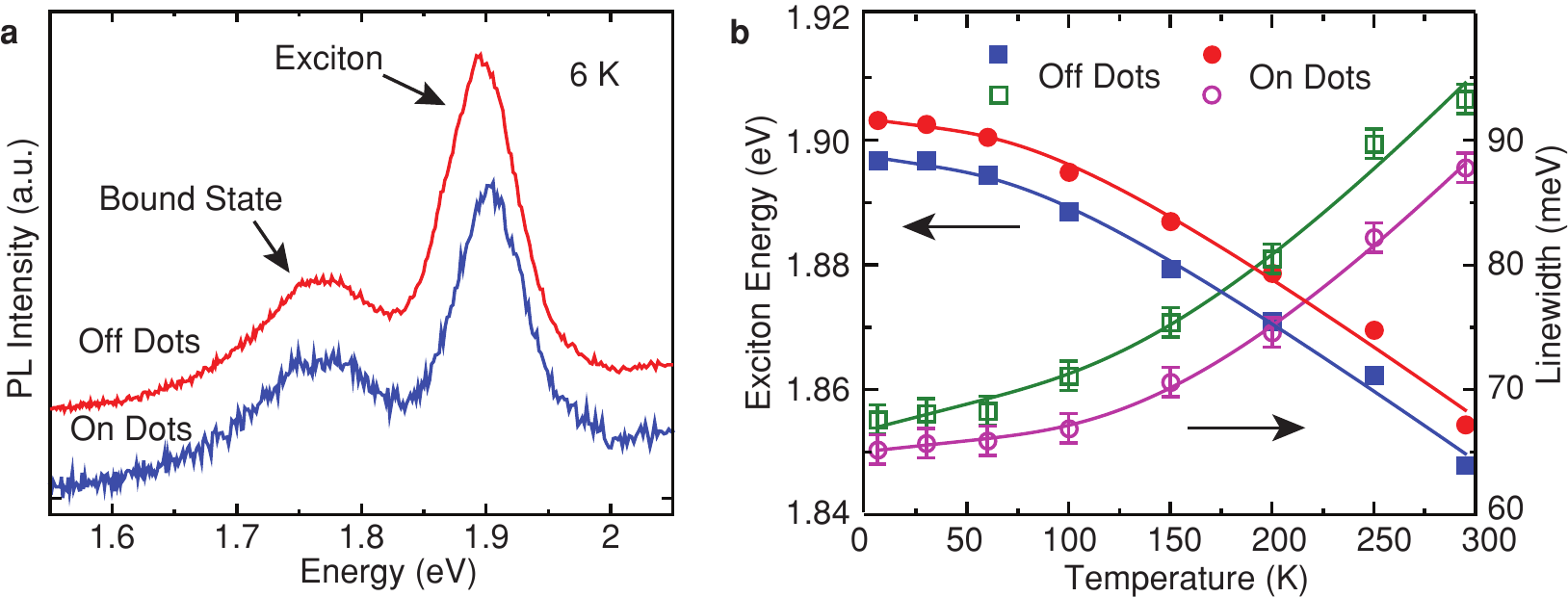}% Here is how to import EPS art
\caption{\textbf{Temperature dependent exciton energy and PL linewidth.} \textbf{a},~PL spectra of the QDs (blue) and continuous monolayer sheet (red) at 6~K both show features of excitonic emission and low-energy bound states emission. \textbf{b},~Temperature dependence of the exciton energy $E_{\textrm{ex}}$ (solid)  and linewidth $\Gamma$ (empty) for QDs (circles) and monolayer (squares). The lines are fits to the M-W equation and equation~(\ref{eq:gamma}), respectively.}
\label{fig:Temperature}
\end{figure*}

\section*{Temperature-dependence of monolayer MoS$_2$ QD emission }

To better understand quantum confinement effects in the monolayer QDs, we perform temperature-dependent PL measurements. Fig.~\ref{fig:Temperature}a shows  PL spectra collected when the excitation is either on the dots or off the dots (on the continuous monolayer). The PL spectra show an exciton peak at about 1.9 eV and a low-energy bound state at 1.77 eV. Monolayer MoS$_{2}$ is well known to exhibit lower-energy bound state luminescence at low temperatures that is not evident at room temperature ~\cite{korn2011low,tongay2013defects}. The weak quantum confinement in our samples does not modify the energy of this bound state PL, indicating that the bound state is localized with a length scale smaller than $\sim 10$~nm.

Temperature dependences of the exciton energy and linewidth are shown in Fig.~\ref{fig:Temperature}b. The exciton energy for QDs and the continuous monolayer can both be fit by the Manoogian-Woolley~(M-W) equation~\cite{M-W}. The difference between monolayer and QD is independent of temperature, consistent with the exciton shift arising from a size-dependent confinement energy.

The exciton linewidth in semiconductors is dominated by inhomogeneity of the material and homogeneous exciton-phonon broadening~\cite{rudin1990temperature}. The temperature dependence of the emission linewidth $\Gamma$ for excitons in a continuous dispersion band can be expressed as
\begin{equation}
\Gamma(T) = \Gamma_{\textrm{inh}}+\gamma_{AC}T+\frac{\gamma_{LO}}{e^{E_{LO}/k_{B}T}-1}
\label{eq:gamma}
\end{equation}
where $\Gamma_{\textrm{inh}}$ is the inhomogeneous linewidth, $\gamma_{AC}$ is the exciton-acoustic (AC) phonon coupling strength, $\gamma_{LO}$ is the exciton-longitudinal optical (LO) phonon coupling strength and $E_{LO}$ is the LO phonon energy~\cite{rudin1990temperature,zhao2002effect}. For monolayer semiconductors, impurities and vacancies in the crystal and variation of the substrate and surface cause significant inhomogeneous broadening $\Gamma_{\textrm{inh}}$. Because the 2D exciton Bohr radius in monolayer MoS$_2$ is only about 1~nm, the material inhomogeneity is expected to remain significant in our comparatively large QDs. As discussed above, the contribution of the QD size distribution to inhomogeneous broadening is small compared to the overall linewidth. At low temperatures, the homogeneous linewidth is dominated by the exciton-AC phonon interaction through the deformation potential and piezoelectric coupling that scatters excitons to intraband states. At higher temperatures, the exciton-LO phonon interaction scatters excitons to both bound and continuum states, with contribution to the homogeneous linewidth  proportional to LO phonon number $1/[\exp(E_{LO}/k_{B}T)-1]$~\cite{rudin1990temperature,zhao2002effect}. Fig.~\ref{fig:Temperature}b shows the fit of the linewidth of continuous monolayer and QDs to equation~(\ref{eq:gamma}) with the LO phonon energy $E_{LO}$ chosen to be 48~meV~\cite{kaasbjerg2012phonon}. The best fit parameters are $\gamma_{AC} = 0.017 \pm 0.012$
~meV/K, $\gamma_{LO} = 101\pm 18$~meV, $\Gamma_{\textrm{inh}}= 65.0\pm 0.9$~meV for QDs and $\gamma_{AC} = 0.042 \pm 0.011
$~meV/K, $\gamma_{LO} = 86\pm 18$~meV, $\Gamma_{\textrm{inh}}= 66.8\pm 0.9$~meV for the monolayer. The slightly smaller $\Gamma_{\textrm{inh}}$ for QDs indicates that in the absence of phonon interactions, QDs exhibit smaller inhomogeneous broadening despite their size distribution. QD linewidth reduction has also been observed in other weakly quantum confined systems~\cite{zhao2002effect}. Exciton-AC phonon interaction ($\gamma_{AC}$) in QDs is smaller than that of the monolayer, consistent with the theoretical predictions and experimental observations in other semiconductor systems such as GaAs~\cite{bockelmann1990phonon,zhao2002effect,borri1999well}. Quantum confinement is known both to enhance and to reduce exciton-LO phonon interactions ($\gamma_{LO}$) in different systems depending on the crystal structure and chemical composition~\cite{pelekanos1992quasi,zhao2002effect,heitz1999enhanced}. In monolayer MoS$_{2}$ QDs, we observe somewhat larger $\gamma_{LO}$. First principle calculations for monolayer MoS$_{2}$ show that the exciton-LO phonon coupling strength $\gamma_{LO}$ is less than 100~meV~\cite{kaasbjerg2012phonon}, which is consistent with our measurement.

\vspace{-5pt}
\section*{Valley polarization in monolayer MoS$_{2}$ QDs}
\vspace{-4pt}

\begin{figure*}
\includegraphics{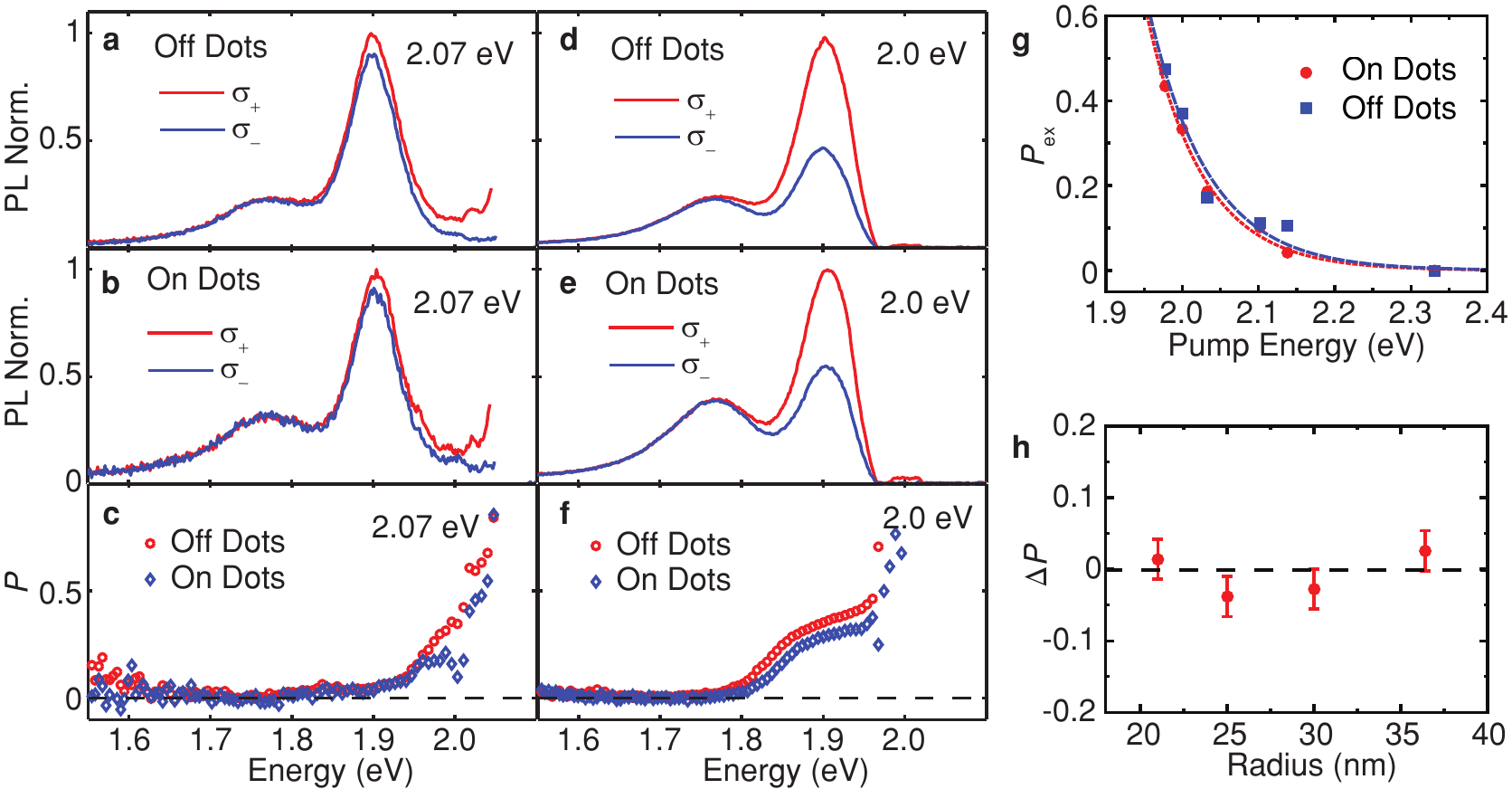}% Here is how to import EPS art
\caption{\textbf{Circularly-polarized emission from monolayer semiconductor QDs.} \textbf{a},\textbf{b}, Circularly-polarized PL for continuous monolayer (off dots) and QDs (on dots) with pump energy of 2.07~eV. \textbf{c}, Polarization $P$ of QDs and monolayer PL with pump energy of 2.07~eV. \textbf{d},\textbf{e}, Circularly polarized PL for continuous monolayer and QDs with pump energy of 2.00 eV. \textbf{f}, Polarization $P$ of continuous monolayer and QDs with pump energy 2.00~eV. \textbf{g}, Pump energy dependence of the continuous monolayer (off dots) and QDs (on dots). The decay is fit to an exponential with  a characteristic energy scale of $\xi = 74$~meV for QDs (red) and $\xi = 79$~meV for a continuous monolayer (blue). \textbf{h}, The negligible change in PL circular polarization with confinement is independent of QD size. Polarization is measured with pump energy at 2.00~eV for \textbf{g} and \textbf{h}. The PL reported in this figure is collected at 6~K. }
\label{fig:polarization}
\end{figure*}
The spin-valley coupling and inversion asymmetry in the TMD monolayer band structure allow selective optical excitation of different valleys using circularly polarized light, which can be detected as circularly-polarized PL emission ~\cite{xiao2012coupled,mak2012control,zeng2012valley}. Valley hybridization as a result of intervalley scattering can reduce emission polarization, as would be expected at elevated temperatures or near monolayer edges~\cite{zeng2012valley,lagarde2014carrier}. To investigate the impact of quantum confinement on valley hybridization in MoS$_2$ QDs, with the associated decrease in emission polarization, we perform pump-energy dependent polarized PL measurements. Fig.~\ref{fig:polarization} shows polarized PL measured on and off QDs with $R = 25$~nm using two different pump energies, 2.07~eV and 2.00~eV. The spectrally-resolved PL polarization is defined by
\vspace{-0pt} \begin{equation}
P = \frac{I_{+}-I_{-}}{I_{+}+I_{-}}
\label{eq:pol}
\end{equation} \vspace{-0pt}
where $I_{+}$ and $I_{-}$ are the right ($\sigma$$_{+}$) and left ($\sigma$$_{-}$) circular polarization-resolved PL intensities. The QD emission exhibits similar polarization to that of the monolayer for each pump energy as shown in Fig.~\ref{fig:polarization}c,f. The slightly smaller polarization for QD emission is due to the higher relative intensity of the tail from the unpolarized low-energy bound state at 1.77~eV and the background. After subtracting a fit to the bound state PL and background, we observe almost identical (within our measurement accuracy) pump-dependent polarization of QD and monolayer emission as shown in Fig.~\ref{fig:polarization}g, where the polarization is calculated similar to equation~(\ref{eq:pol}) but using the integrated intensities of the exciton PL peak.  For both QDs and monolayer, the polarization increases for pump energy closer to the exciton resonance. The pump energy dependence shown in Fig.~\ref{fig:polarization}g can be characterized by a phenomenological exponential curve  $P \sim e^{-(E_{\textrm{pump}}-E_{\textrm{ex}})/\xi}$, where $E_{\textrm{pump}}$ is the pump photon energy, $E_{\textrm{ex}}$ is the exciton energy, and $\xi$ describes the energy decay scale. The fit to QDs (monolayer) gives $E_{\textrm{ex}}$ of $1.916\pm0.006$ ($1.918\pm0.014$)~eV and $\xi$ of $0.074\pm0.006$ ($0.079\pm0.013$)~eV. The parameters for QDs and monolayers are consistent with each other, and the temperature dependence of $P$ is similar to that previously reported for MoS$_2$ and associated with acoustic phonon-driven intervalley scattering~\cite{kioseoglou2012valley}.  Fig.~\ref{fig:polarization}h shows the difference of emission polarization between QD and continuous monolayer sheet for varying QD size. Within the measurement accuracy, nearly unchanged polarization is observed across all the QD sizes.

Valley structure in low-dimensional materials is known to be perturbed by confinement in graphene and silicon QDs~\cite{ goswami2007controllable,saraiva2009physical,trauzettel2007spin}. The persistence of valley polarization in monolayer TMD quantum dots can be explained by both energy and wavefunction arguments~\cite{liu2014intervalley}.  For a patterned QD, the confinement potential is spin independent and intervalley scattering occurs between the same spin states. Because of the large valence band spin splitting of 150~meV in monolayer MoS$_{2}$~\cite{xiao2012coupled,liu2013three}, intervalley scattering of holes is suppressed. Despite the strong confinement potential dependence of intervalley scattering of the nearly-degenerate conduction band, the upper bound of the intervalley coupling in monolayer MoS$_{2}$ QDs is 0.1~meV for small size QDs (20~nm in diameter) with vertical wall confinement potentials~\cite{liu2014intervalley}. This intervalley coupling is negligible compared to the energy difference between the same spin states in different valleys, which is several meV in MoS$_{2}$~\cite{liu2013three}, and even larger in other TMD materials. Unlike in graphene and silicon QDs, valley hybridization is additionally suppressed in TMD QDs because the exciton wavefunction vanishes at the boundaries where intervalley coupling is significant~\cite{liu2014intervalley}. Since the intervalley scattering strength is small, valley polarization is expected to persist in our QDs. Our observation of valley-polarized emission in quantum confined monolayer QDs confirms the prediction that intervalley scattering is negligible in monolayer MoS$_{2}$ QDs with radii as low as 20~nm and that the valley index is a good quantum number in this confinement regime.

\section*{Conclusion}
The discovery of the direct bandgap in monolayer TMD semiconductors~\cite{mak2010atomically} has prompted the search for layer-dependent low-dimensional analogs of traditional semiconductor devices such as heterostructures~\cite{ Geim2013, Cheng2014, Yu2013, Novoselov2012}. The controlled size-dependence of low-dimensional excitons is an important consequence of quantum confinement useful for many applications of semiconductor quantum dot materials. Our nanofabrication approach enables size-tunable quantum confinement in monolayer MoS$_{2}$ QDs using a controllable top-down semiconductor-compatible fabrication process. Distinct from recent reports of localized quantum emitter defects in monolayer WSe$_{2}$~\cite{he2015single,srivastava2015optically,koperski2015single, chakraborty2015voltage}, our results demonstrate size-dependent quantum confinement of delocalized band carriers. The emission energy shifts are limited by the tightly-bound exciton in monolayer TMDs.

Unlike in many low-dimensional systems such as silicon and graphene QDs~\cite{goswami2007controllable,saraiva2009physical,trauzettel2007spin}, the carrier valley polarization persists in monolayer MoS$_{2}$ QDs due to negligible intervalley scattering. The inheritance of monolayer valley properties in monolayer semiconductor QDs with size-controlled exciton energies suggests that TMD monolayer QDs are a potential platform for harnessing the valley degree of freedom of a single electron~\cite{liu2014intervalley,kormanyos2014spin}. Extending semiconductor-compatible nanofabrication of monolayer QDs to achieve isolated single emitters with non-trivial band symmetry can help to realize multimodal building blocks for integrated quantum information systems based on spin and valley.

\section*{Methods}

\textbf{Sample preparation.} Monolayer MoS$_{2}$ flakes are obtained from a bulk crystal (SPI Supplies) using mechanical exfoliation. The flakes are dry transferred onto a SiO$_{2}$/Si substrate with pre-written alignment marks~\cite{wei2015silicon}. Identification of monolayer MoS$_{2}$ is done using atomic force microscopy (AFM), optical contrast, and PL spectroscopy as described in detail previously~\cite{wei2015silicon}. Samples are annealed in an Ar/H$_{2}$ environment at 400~C before patterning with electron beam (e-beam) lithography and reactive ion etching using CHF$_3$/O$_2$ gas. The resist is not removed at the end of processing to prevent destruction of the monolayer QDs. Additional details of the sample fabrication can be found in the Supplementary Information.\\

\textbf{AFM characterization.} QD size is characterized by atomic force microscopy (AFM) in noncontact mode. The area of each QD is measured through image processing. To obtain the QD size distribution, the somewhat non-circular QD shapes are modelled as circular dots with diameter $D$, so that $\pi D^2/4$ equals to the area of each dot from the AFM scan.\\

\textbf{Optical measurement.} Standard PL characterization was performed with a 532~nm semiconductor laser, and near-resonant circularly-polarized PL is measured with a tunable pulsed optical parametric oscillator whose linewidth was narrowed to approximately 1~nm using a double-grating system. The pump laser is focused onto the sample through a 100$\times$ long-working distance objective with numerical aperture of 0.65 and a spot size $\sim$ 1~$\mu$m. The pump power is kept less than 40~$\mu$W to suppress heating effects during low temperature measurements. The objective can be spatially scanned across the sample with sub-micron resolution. The sample is mounted in a closed-cycle cryostat capable of reaching temperatures between 4~K and 350~K. The PL is collected through the same objective and analyzed by a spectrometer with a CCD detector. Energy and linewidth shifts for QDs are reported relative to the continuous monolayer control sample from the same exfoliated flake. Reported best-fit parameters are obtained from weighted least-squares fitting with parameter uncertainties estimated using resampling. For circularly-polarized PL, the samples are stabilized at 6~K. The pump polarization is controlled using a linear polarizer and a liquid crystal retarder.  The emitted PL is analyzed using a quarter waveplate and a linear polarizer before being coupled into the spectrometer.  The polarization of the pump laser is measured to be greater than 99.9\% with this apparatus.

\section*{Acknowledgements}

This work is supported by the Institute for Sustainability and Energy at Northwestern (QD device fabrication), the U.S. Department of Energy, Office of Basic Energy Sciences, Division of Materials Sciences and Engineering (DE-SC0012130) (spectroscopy), and Argonne National Laboratory. Sample preparation and characterization was partially supported (E.J.L.) by the National Science Foundation’s MRSEC program (DMR-1121262) and made use of its Shared Facilities at the Materials Research Center of Northwestern University.  Use of the Center for Nanoscale Materials was supported by the U. S. Department of Energy, Office of Science, Office of Basic Energy Sciences, under Contract No. DE-AC02-06CH11357. The work utilized the Northwestern University Micro/Nano Fabrication Facility (NUFAB) and the EPIC facility (NUANCE Center-Northwestern). The authors acknowledge the assistance of Venkat Chandrasekhar for equipment access. N.P.S. is an Alfred P. Sloan Research Fellow.

\bibliography{Wei_2015}

\end{document}